\pdfoutput=1
\documentclass[conference]{IEEEtran}
\usepackage{cite}
\usepackage{graphicx, amsmath, fixmath, caption, amssymb, tabu}
\usepackage[caption=false]{subfig}
\usepackage{commath}
\usepackage{bm}
\usepackage{verbatim}
\usepackage{svg}

\graphicspath{ {images/} }
\ifCLASSINFOpdf
\else
\fi
\newcommand{\quotes}[1]{``#1''}

\begin{document}
%
\title{\LARGE \bf Fuzzy Based Wellness Analyzer\\[-3.0ex]}



%
\author{\IEEEauthorblockN{Rohit Ner, Dibya Prakash Das, Rishabh Kumar, Shubhika Garg}
\IEEEauthorblockA{Department of Mathematics, Indian Institute of Technology Kharagpur, India\\
\tt \{rohitner, dibyadas, rishabhkumar, shubhika.garg\}@iitkgp.ac.in
}
}


\maketitle

\begin{abstract}
Social health and emotional wellness is a matter of concern in today\rq s urban world. Being the part of a metropolis has an effect on mental health through the influence of increased stressors and factors such as overcrowded and polluted environment, high levels of violence, and reduced social support. It is important to realize that only healthy citizens can constitute together a smart city. In this paper, we present a fuzzy-based approach for analyzing the well being of a person. We track the general day to day activities of a person and analyze its performance. To do so, we divide the factors affecting the wellness of a person into three components which are the physical, productive and social. Using these parameters, we output a coefficient for the overall well being of a person.
\end{abstract}

\begin{IEEEkeywords}
Fuzzy Logic, Public health, Smart cities
\end{IEEEkeywords}

%
\IEEEpeerreviewmaketitle

\section{Introduction}

Over half of the world\rq s population lives in urban areas. Urbanization presents opportunities and risks, as well as enormous challenges for maintaining and improving human health and well being. Degrading wellness is proving to be an important issue in the modern world. Cases of suicide and depression have been rising over the years. One of the main barriers in handling these situations is the improper reporting of the problem by the victim and their lack of communication with the concerned close ones. In recent times several such issues have been addressed by researchers using fuzzy logic. Guntuku et al. \cite{Guntuku} identify mentally ill users by patterns in their language and online activity and propose automated analysis by large-scale passive monitoring. Kelley et al. \cite{Kelley} discuss the challenges and opportunities in leveraging self-tracking for mental wellness. A fuzzy optimization model based on chanced constrained optimization to give health related recommendations is proposed by Mezei et al. \cite{Mezei}. Liu et al. \cite{Liu} examine the correlation between depression and speech using acoustic features of patients. Lutz et al. \cite{Lutz} propose a framework that uses fuzzy-set theory to measure human well-being in consistence with Sen’s Capability Approach. 

From the review, it has been seen that not many references address the problem of online wellness tracking using fuzzy logic. We introduce a novel approach to analyze the well being of a person by his tracking the daily activities. We investigate the common symptoms of a depressed person which include a loss of interest in daily activities, loss of appetite, change in the sleep cycle, change in health and lack of social interactions. Our approach compensates for the inability to define ideal health, social and work conditions due to their subjective nature by using a fuzzy inference mechanism.

\subsection{About Fuzzy Logic}

Fuzzy logic is used widely in all domains since the past few decades. Its cross-platform utility since its inception is one of the primary reasons for such instantaneous growth. Fuzzy logic also sometimes called as grey logic discards the principles of Propositional Logic which revolves around two truth values and instead proposes a new theory of \lq Partial Truth\rq, also referred as the degree of membership as per the suggestions of Zadeh \cite{ZADEH1965338}.

Let $\mathbf{S}$ be a non empty set, called the \textit{universe set}. Now, consider any crisp set $A \subset \mathbf{S}$. A characteristic function $\mathbold{\chi}_A$ is defined as

$$
\mathbold{\chi}_A(x) = 
\begin{cases}
    1, & \text{if } x\in A\\
    0, & \text{otherwise}
\end{cases}
$$

$\mathbold{\chi}_A$ maps every element of $\mathbf{S}$ to a value of either $1$ or $0$. Let $B$ be a fuzzy set such that $B \subset \mathbf{S}$. For any $x \in S$ the function $\mathbold{\mu}_B(x)$  defined as $\mathbold{\mu}_B:\mathbf{S}\rightarrow[0, 1]$ yields the degree of belonging of $x$ to the set $B$ ranging between $0$ and $1$. $\mathbold{\mu}_B(x)$ is known as the membership function. The fuzzy set\cite{Klir:1996:FSF:234347} theory provides a new dimension which is the limitation of classical theory and helps us to express uncertainty in a bold manner.

\subsection{Problem Formulation}

The problem can be divided into three major parts:

\begin{itemize}
\item \textbf{Context Recognition}: Collecting sensor measurements and labels describing the user's context through various methods including self-reporting.
\item \textbf{Base Level Inference}: Fuzzify the crisp input labels and calculate the values membership functions of broadly categorized components.
\item \textbf{Top Level Inference}: The components are defuzzified and fed into the top controller to get a output the level of percentage well being of the person.
\end{itemize}

\begin{figure}[h!]
\centering
\captionsetup{justification=centering}
\noindent \includegraphics[width=0.9\linewidth]{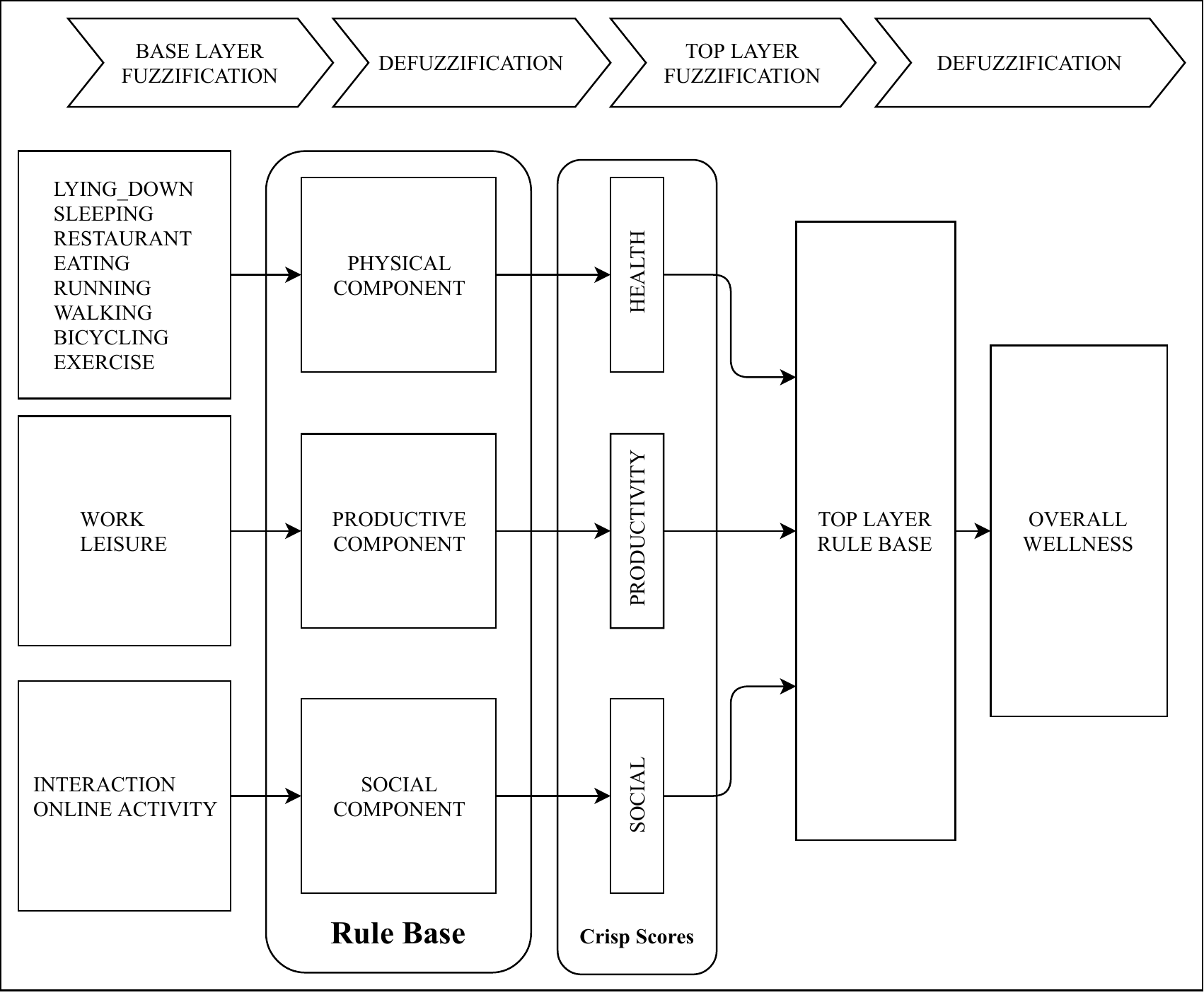}
\caption{Fuzzy logic system architecture for determining a person\rq s well being.}
\label{architecture}
\end{figure}

\section{Working Principle}

The general work flow of this paper is shown in Figure \ref{architecture}.

\subsection{Context Recognition}

Context recognition is used to identify the user\rq s real-time contextual information from data of sensors, using relevant signal processing approaches. There are several approaches mentioned in the literature that incorporate stimuli from smartphone sensors which include audio-based context recognition \cite{Choi}, involving dynamics-based using smart-watches \cite{Faye}. Most of the approaches need data that is collected under strictly constrained conditions with the investigator instructing subjects to execute fabricated chores. There is a need for an approach capable of performing given the practical in-the-wild conditions. Vaizman et al. \cite{Vaizman} introduce an approach which recognizes the context on the basis of natural behavioral content. The system consists of single-sensor classifiers and sensor-fusion. We use this approach for our needs.

On investigation, the common symptoms of depression turn out to be a loss of interest in daily activities, appetite loss, altered sleep cycle and lack of social interaction. We divided the factors affecting the overall wellness of the person into three components namely physical, productive and social. A person in an urban area carries one\rq s smartphone almost everywhere. We use the ExtraSensory App \cite{ExtraSensory} for the purpose of collecting the data using the in-the-wild approach. The application provides a self-reporting interface through which users can report labels signifying their current activity.

\begin{table}[ht]
\small
\captionof{table}{Factors affecting physical component}
\label{tab_physical}
\begin{center}
\def\arraystretch{1.7}
\begin{tabular}{|l|l|}
\hline
 \bf Factor & \bf Label \\
 \hline
 Sleep & \texttt{LYING\_DOWN}, \texttt{SLEEPING} \\
 \hline
 Diet & \texttt{RESTAURANT}, \texttt{EATING} \\
 \hline
 Exercise & \texttt{RUNNING}, \texttt{WALKING}, \texttt{BICYCLING}, \texttt{GYM},\vspace{-5pt} \\
 & \texttt{EXERCISE}\\
 \hline
\end{tabular}
\end{center}
\end{table}

\begin{table}[ht]
\small
\captionof{table}{Factors affecting productive component}
\label{tab_productive}
\begin{center}
\def\arraystretch{1.7}
\begin{tabular}{|l|l|}
\hline
 \bf Factor & \bf Label \\
 \hline
 Work & \texttt{IN\_CLASS}, \texttt{IN\_A\_MEETING}, \texttt{AT\_SCHOOL},\vspace{-5pt}
 \\ & \texttt{LOC\_main\_workplace} \\
 \hline
 Leisure & \texttt{WATCHING\_TV}, \texttt{SINGING}, \texttt{SHOPPING} \\
 \hline
\end{tabular}
\end{center}
\end{table}

\begin{table}[ht]
\small
\captionof{table}{Factors affecting social component}
\label{tab_social}
\begin{center}
\def\arraystretch{1.7}
\begin{tabular}{|l|l|}
\hline
 \bf Factor & \bf Label \\
 \hline
 Interaction & \texttt{WITH\_FRIENDS}, \texttt{WITH\_CO-WORKERS},\vspace{-5pt}
 \\ & \texttt{TALKING}, \texttt{AT\_A\_PARTY} \\
 \hline
 Online Activity & \texttt{SURFING\_THE\_INTERNET},\vspace{-5pt} \\&\texttt{PHONE\_IN\_HAND}\\
 \hline
\end{tabular}
\end{center}
\end{table}

\subsubsection{Physical}

Physical health ensures that the body is functioning properly. One must gain fiber, use one\rq s parts and maintain an amenable environment for stable growth. The physical health of a person is analyzed by subdividing the component into three categories \textit{sleep}, \textit{diet} and \textit{exercise}. Table \ref{tab_physical} shows the labels in the application used to determine the quantity of these categories.

\subsubsection{Productive}

The productivity of a person depends on several factors like the capacity of the person to reach a certain level of efficiency by putting in energy in the form of manual or cognitive activity. According to the findings of Mokana et al. \cite{Mokana}, work-life balance was found to be significantly dependent on work culture and technological improvements. An appropriate balance between the work time and the leisure time is suggested for good health. We hence choose the categories of \textit{work} and \textit{leisure} as the primarily affecting factors of productivity. Table \ref{tab_productive} shows the labels used for determining the amount of workload and leisure time spent by the person.

\subsubsection{Social}

Social wellness is the positive effect relationships have on one\rq s mental and physical health. Quantitatively, it is a measure of health based on the number of close and personal friends one has as well as how often one spends time with them in person. Online social networking has become a part of health promoting interventions. Peer pressure, social support and data exchange in these virtual communities are seen to affect health behaviors. We divide the categories affecting this component as \textit{interaction} and \textit{online} activity. The corresponding labels are shown in Table \ref{tab_social}.

\subsection{Base Level Inference}

\begin{figure}[b]
\centering
\captionsetup{justification=centering}
\includegraphics[width=\linewidth]{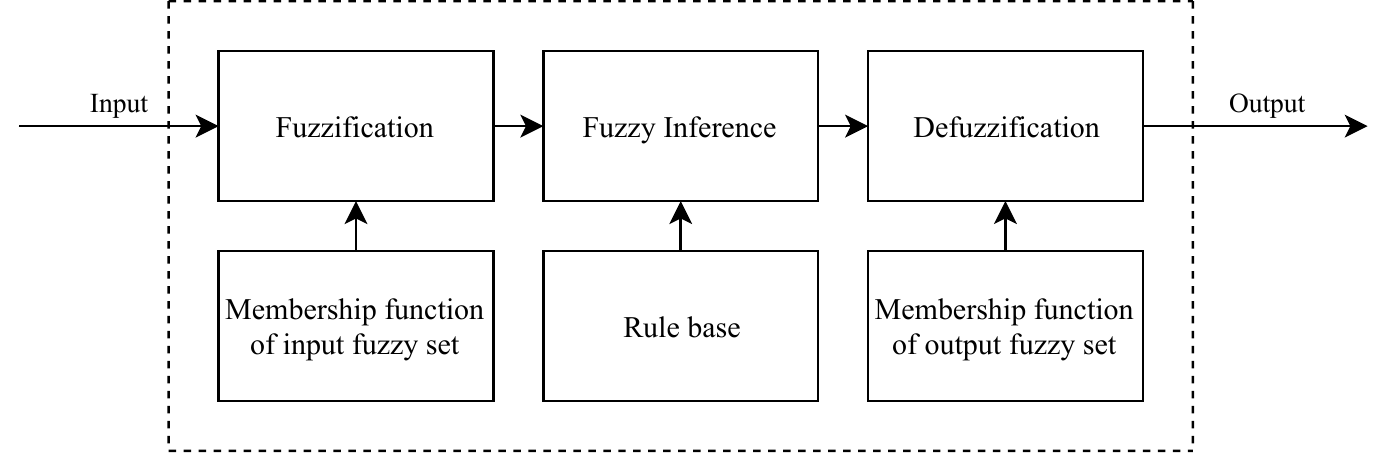}
\caption{General architecture of a fuzzy controller}
\label{architecture_controller}
\end{figure}

Fuzzy inference is the process of mapping from a given input to an output using fuzzy logic. This system helps to model linguistic variables in terms of a well-established quantitative input and providing a crisp output. Multiple fuzzy control models exist in theory. The general architecture is shown in Figure \ref{architecture_controller}. The two most commonly used classes of fuzzy controllers are Mamdani controller and Sugeno controller.

\textbf{Mamdani Controller} The Mamdani controller \cite{Mamdani} is a universal fuzzy controller i.e., it can take the form of any non-linearized controller depending on the membership functions. Mamdani controller is generally used to convert linguistic variables to a fuzzy $ IF-THEN $ rule-base using $ t-norm $ as $ min(x) $ and $ t-conorm $ as $ max(x) $ as per the antecedent to compute the take logical decisions on the antecedents. Then the output is converted to a defuzzified value for producing the final score using an optimal defuzzification method like center-of-gravity defuzzifier. Multiple research shows that Mamdani controller is well suited for human inputs than the Sugeno counterpart. Although the later is more computationally efficient, Mamdani controller is much more intuitive and simplifies rule base formulation.

For calculating the scores of the three components, the labels of the categories on which each component depends are fed into the base-level inference system. The membership functions obtained as outputs are then defuzzified using the center-of-gravity criterion. The defuzzified output provides the percentage of the respective component exhibited by the person.

\subsection{Top Level Inference}

The crisp values of the three components are fed into the subsequent controller to obtain the membership function of the overall wellness of the person which is then converted to a percentage using the center-of-gravity defuzzification.

\section{Experiment}

As stated earlier, we test our algorithm on data that is in its organic form and obtained in the natural conditions of urban living. We use data of 60 users that was obtained by Vaizman et al. \cite{ExtraSensory}. Every user was allotted a universally unique identifier (UUID). The dataset has thousands of examples, typically taken in approximate intervals of 1 minute for every user. The measurements were obtained from the sensors on user\rq s smart-phone and a smart-watch. Self-reported context labels are present in most of the examples. All the relevant context labels could be reported by the user for the immediate future, and it was also possible to choose the amount of time for which the same labels will stay relevant.

\begin{figure}[!b]
\centering
\captionsetup{justification=centering}
\noindent \includegraphics[width=\linewidth]{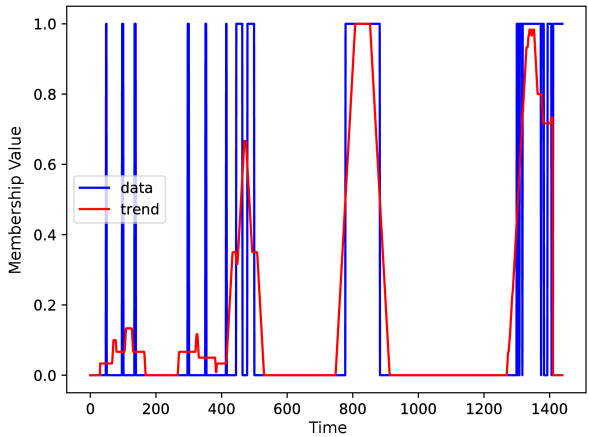}
\caption{Captured trend of the \textit{online} subcomponent using STL decomposition}
\label{stl}
\end{figure}

\begin{table}[ht]
\small
\captionof{table}{Rule base for top level inference with memberships as H=High, M=Medium, L=Low}
\label{rulebase}
\begin{center}
\def\arraystretch{1.7}
\begin{tabular}{|c|c|c|c|}
\hline
 \bf Health & \bf Productive & \bf Social & \bf Overall \\
 \hline
 L & L & L & L \\
 \hline
 M & L & L & L \\
 \hline
 H & L & L & L \\
 \hline
 M & M & L & M \\
 \hline
 H & M & L & M \\
 \hline
 \dots & \dots & \dots & \dots \\
 \hline
 L & M & H & M \\
 \hline
 H & M & H & H \\
 \hline
 L & H & H & M \\
 \hline
 M & H & H & H \\
 \hline
 H & H & H & H \\
 \hline
\end{tabular}
\end{center}
\end{table}

\begin{figure*}
\centering
\subfloat[$\mu_{sleep}$]{%
  \includegraphics[width=0.33\textwidth]{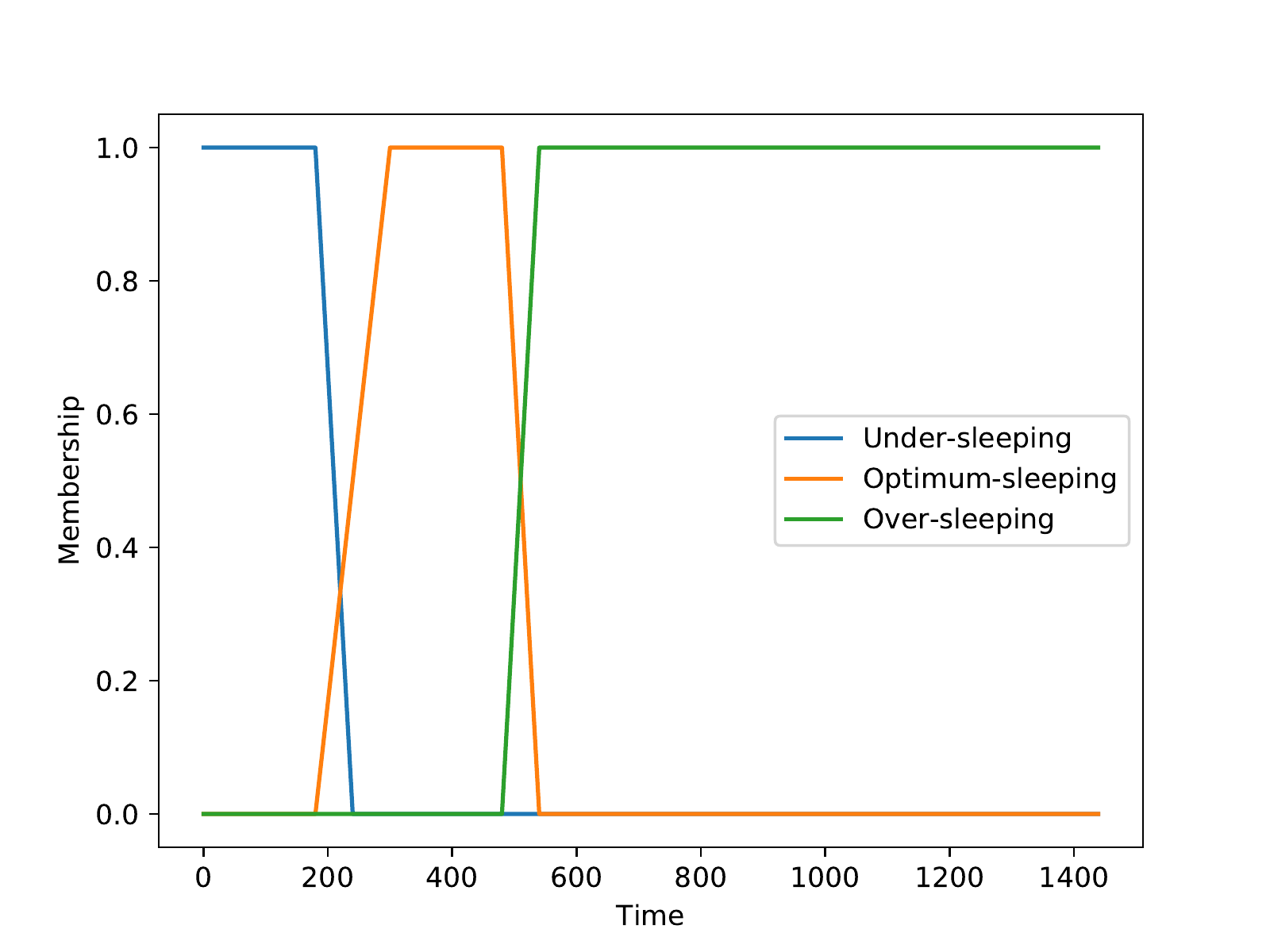}%
}
\subfloat[$\mu_{eat}$]{%
  \includegraphics[width=0.33\textwidth]{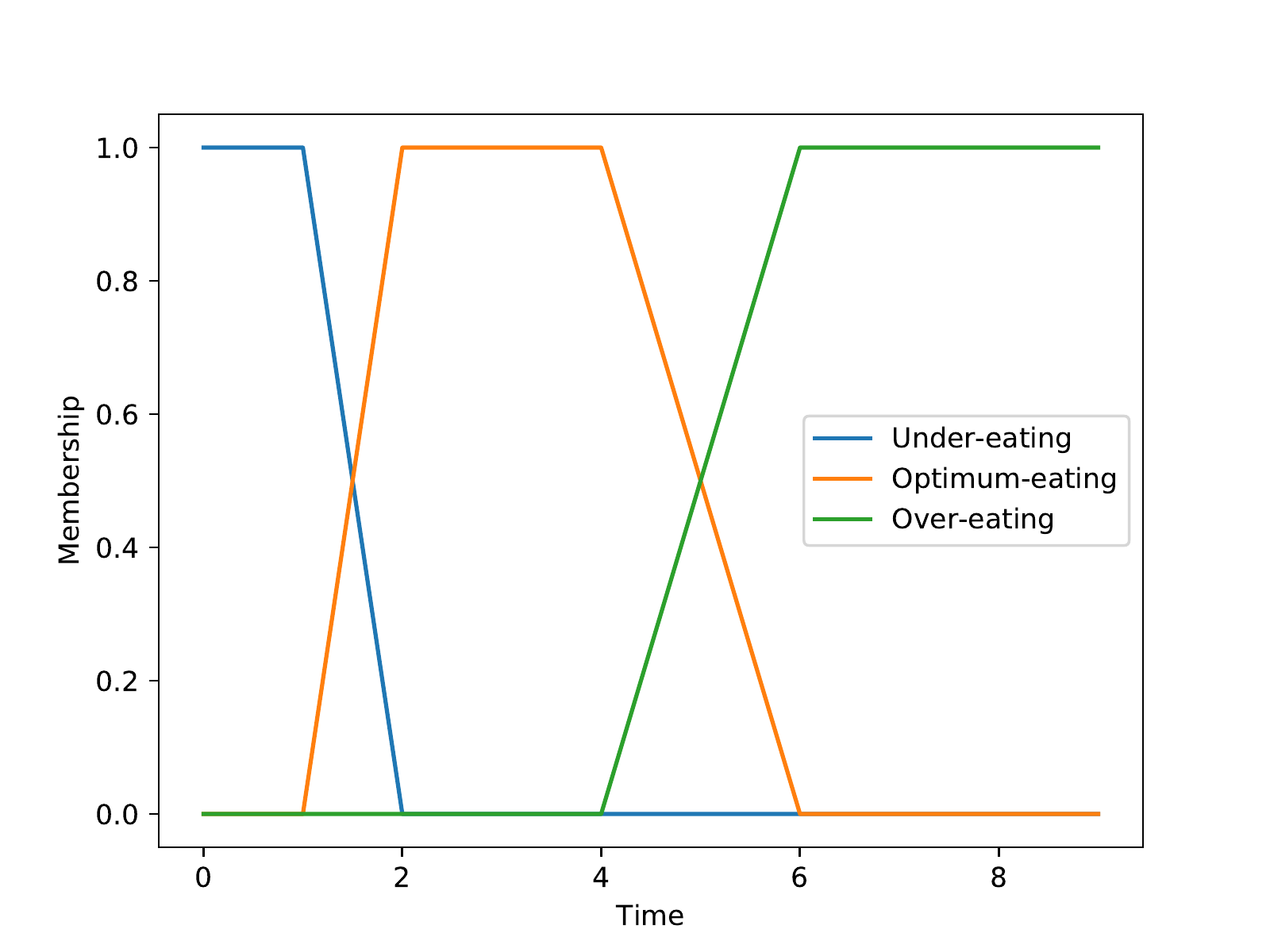}%
}
\subfloat[$\mu_{exercise}$]{%
  \includegraphics[width=0.33\textwidth]{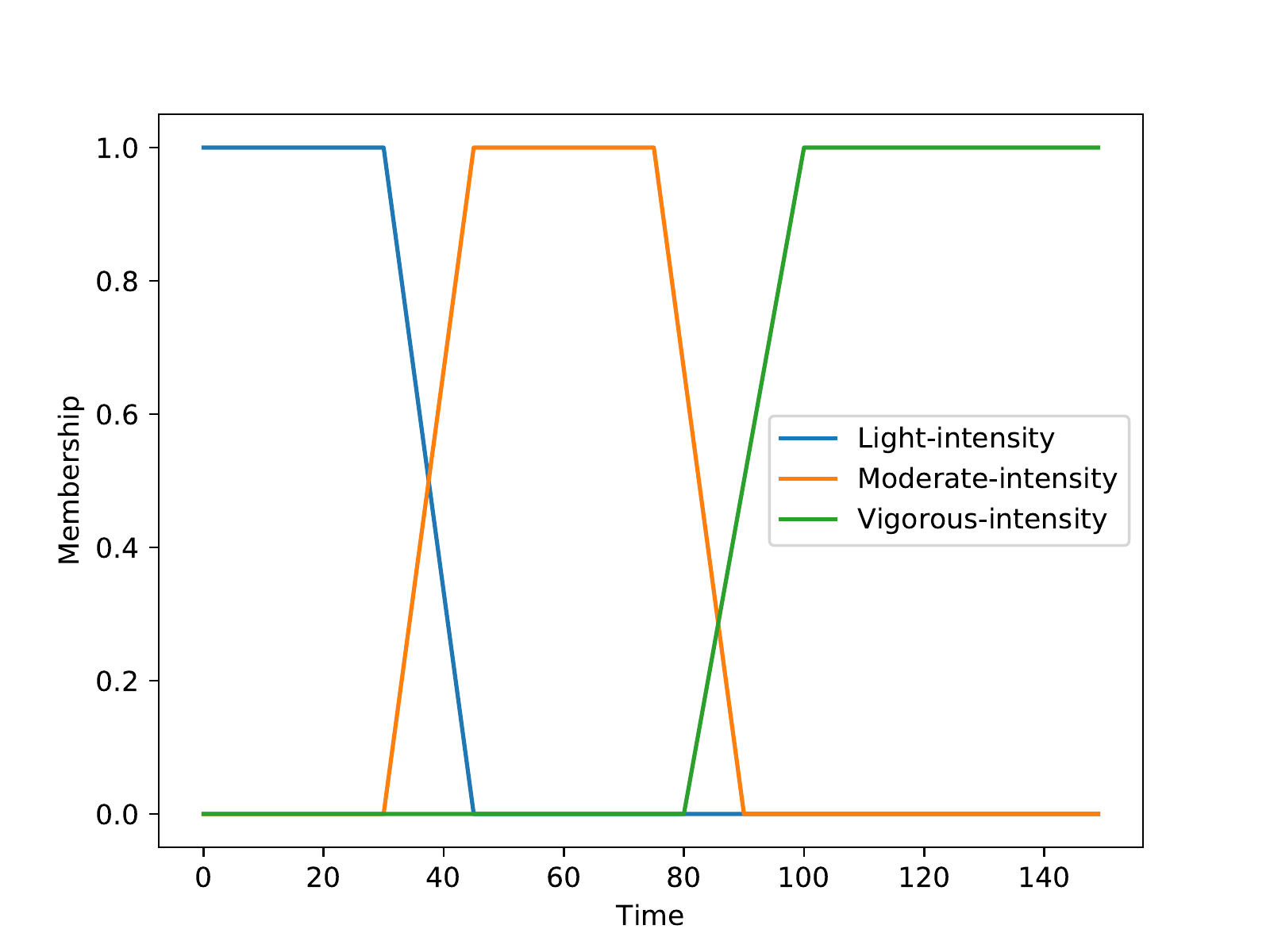}%
}\hfill
\subfloat[$\mu_{work}$]{%
  \includegraphics[width=0.33\textwidth]{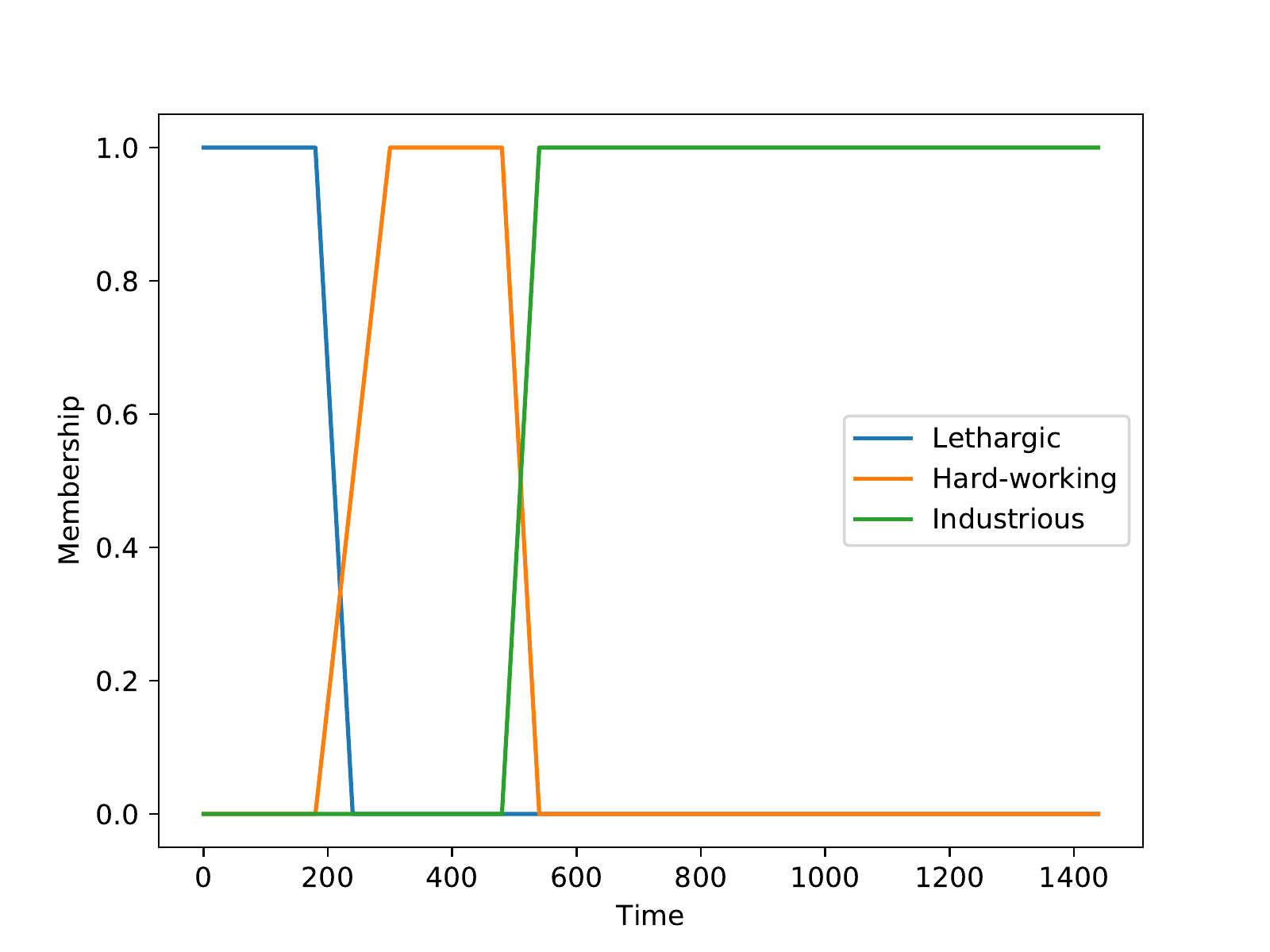}%
}
\subfloat[$\mu_{leisure}$]{%
  \includegraphics[width=0.33\textwidth]{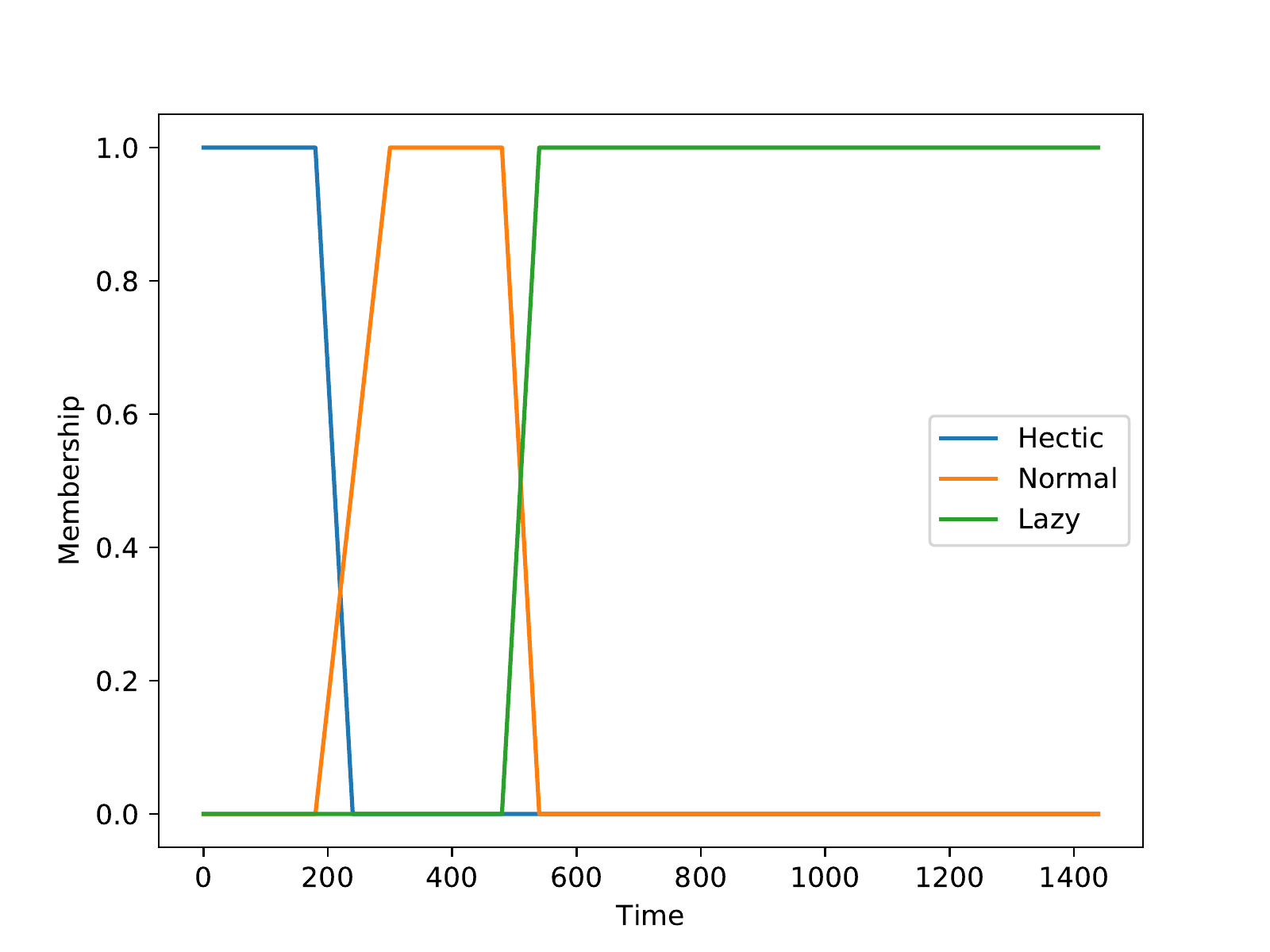}%
}
\subfloat[$\mu_{interaction}$]{%
  \includegraphics[width=0.33\textwidth]{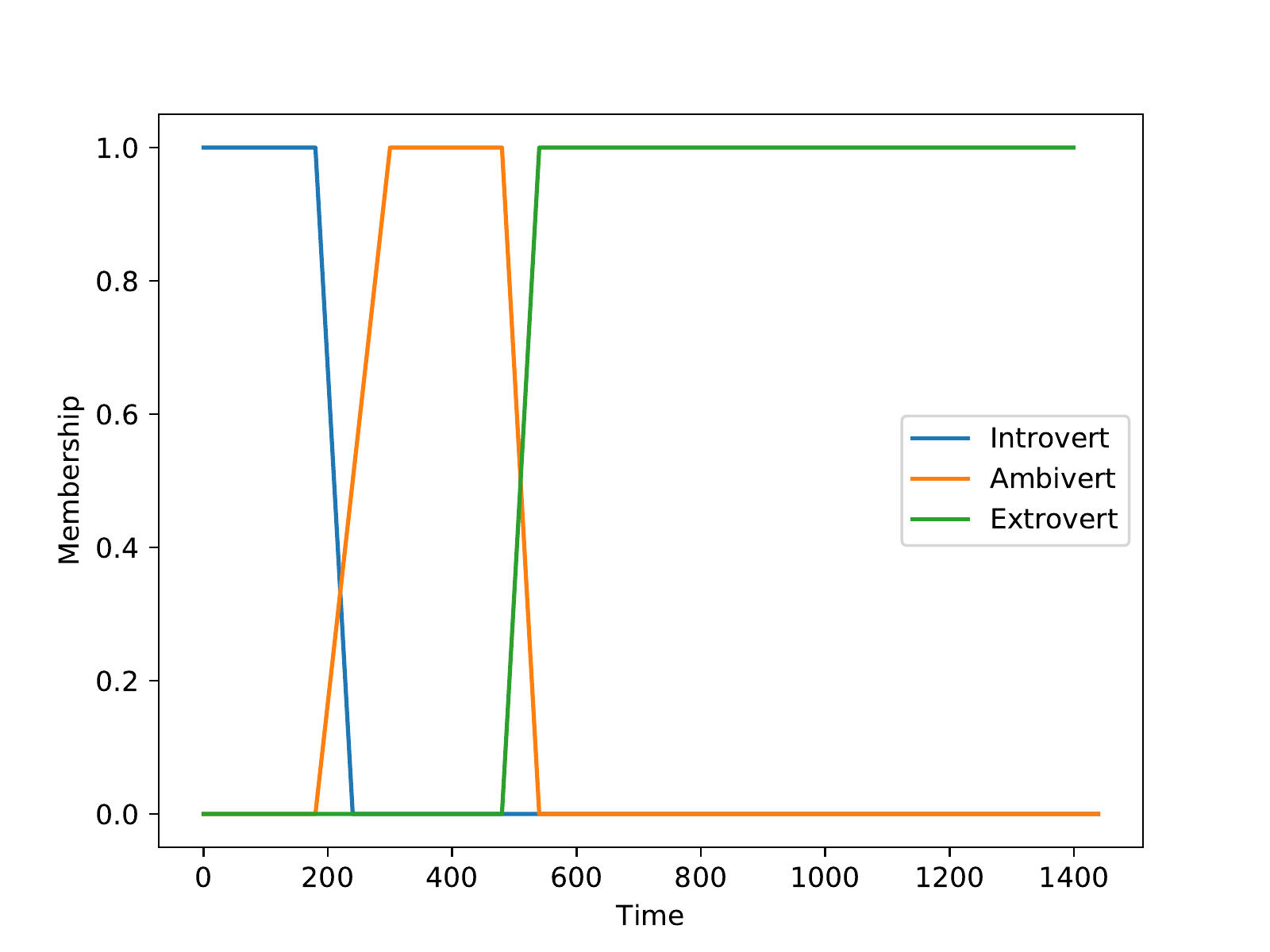}%
}\hfill
\subfloat[$\mu_{online}$]{%
  \includegraphics[width=0.33\textwidth]{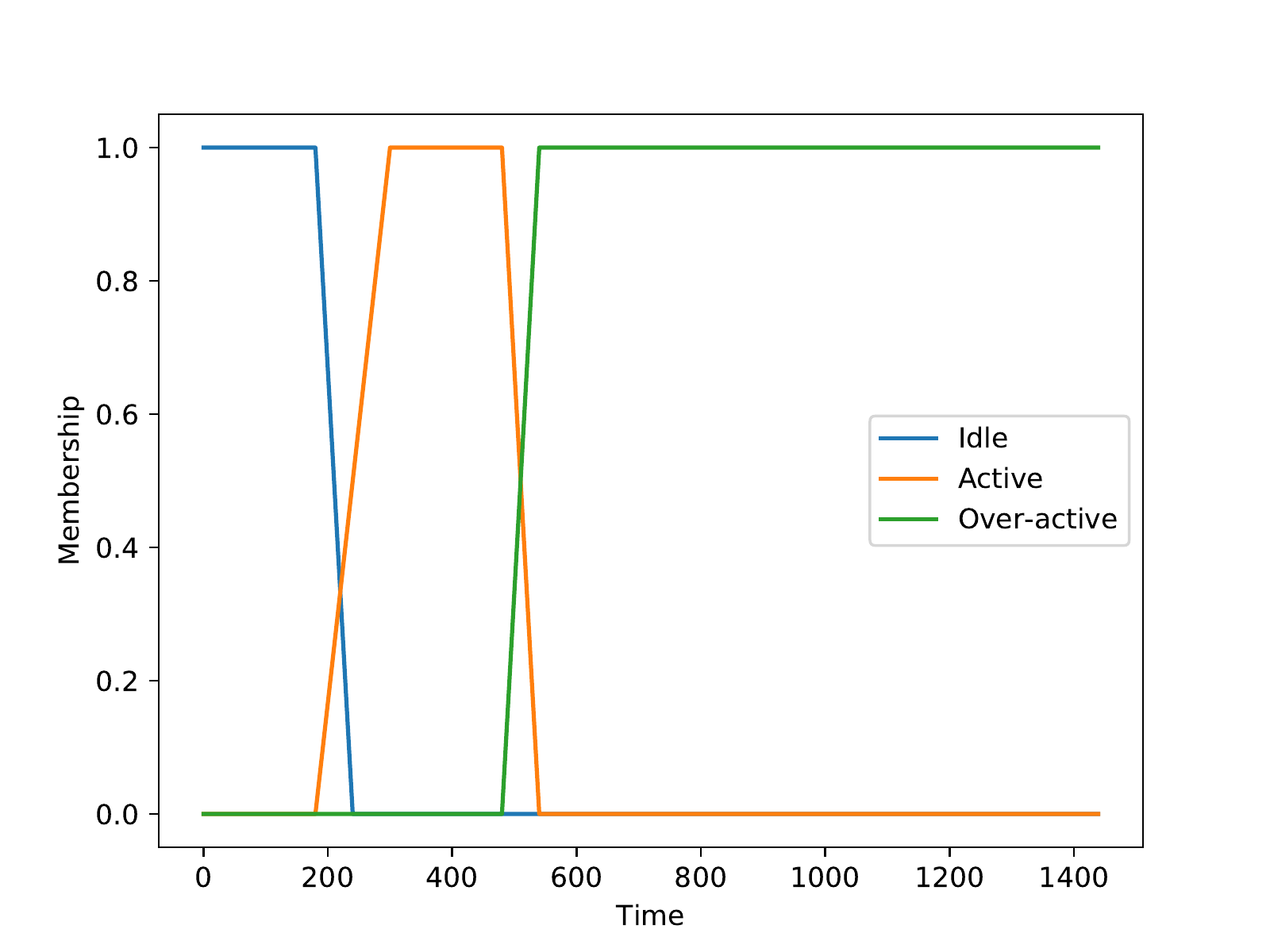}%
}\hfill
\caption{Membership functions for linguistic terms of all categories of base-layer inference input with respect to time.}
\label{memberships}
\end{figure*}

The membership functions of some of the linguistic terms used in the study are shown in Figure \ref{memberships}. The rule base entries were decided after an extensive lookout for the dependence of the three components \textit{physical}, \textit{productive} and \textit{social} on the overall wellness. An excerpt from the rule base for the top layer inference is shown in Table \ref{rulebase}.

\textbf{STL Decomposition} One of the issues faced was the irregularities in the reporting done by the participant. To determine the number of times a person does an activity, we use Seasonal and Trend decomposition using Loess. As seen in Figure \ref{stl}, this method helps in ruling out the noise generated in the data during activity tracking.

\begin{table}[ht]
\small
\captionof{table}{Scores and list of moods of a sample from the dataset.}
\label{results}
\begin{center}
\def\arraystretch{1.7}
\begin{tabular}{|p{0.1\linewidth}|p{0.2\linewidth}|p{0.2\linewidth}|p{0.2\linewidth}|}
\hline
\bf Total & \bf Health & \bf Productive & \bf Social\\
 \hline
 UUID & \multicolumn{3}{l|}{\tt 8023FE1A-D3B0-4E2C-A57A-9321B7FC755F}\vspace{-5pt}\\
 Labels & \multicolumn{3}{l|}{\texttt{HAPPY}, \texttt{ATTENTIVE}, \texttt{DREAMY}}\\
 \hline
 \textbf{82.067} & 81.877 & 48.550 & 48.137 \\
 \hline
 UUID & \multicolumn{3}{l|}{\tt CA820D43-E5E2-42EF-9798-BE56F776370B}\vspace{-5pt}\\
 Labels & \multicolumn{3}{l|}{\texttt{HAPPY}, \texttt{HUNGRY}, \texttt{ATTENTIVE}}\\
 \hline
 \textbf{58.075} & 80.721 & 48.550 & 25.294 \\
 \hline
 UUID & \multicolumn{3}{l|}{\tt 806289BC-AD52-4CC1-806C-0CDB14D65EB6}\vspace{-5pt}\\
 Labels & \multicolumn{3}{l|}{\texttt{STRESSED}, \texttt{TIRED}, \texttt{SLEEPY}}\\
 \hline
 \textbf{86.826} & 81.877 & 48.550 & 81.877 \\
 \hline
 UUID & \multicolumn{3}{l|}{\tt B09E373F-8A54-44C8-895B-0039390B859F}\vspace{-5pt}\\
 Labels & \multicolumn{3}{l|}{\texttt{ACTIVE}, \texttt{CALM}, \texttt{TIRED}}\\
 \hline
 \textbf{50.634} & 81.877 & 52.262 & 20.416 \\
 \hline
\end{tabular}
\end{center}
\end{table}

The ExtraSensory application has an option to select mood labels when self-reporting. For the purpose of evaluation, we aggregate the three most reported labels provided by the user within the time frame of which the observations are considered. Table \ref{results} indicates the similarity between the reported moods and the scores generated by our approach.


%

\section{Concluding Remarks}
A new fuzzy inference based approach is proposed in the context of wellness analysis. The proposed algorithm is simple and easy to implement in practice. The accuracy of the algorithm can be improved by enhancing the passive reporting of the user using Internet-of-Things and sensor integration at public spaces. Hybrid neuro-fuzzy models can be used for better prior estimation. There is a need for a society where one can leverage the connectivity on a social level owing to the respective platforms. The fuzzy-based system discussed here has potential applications in the development of self-awareness and proper circulation of data for an increased mutual understanding.

\section*{Acknowledgment}
We would like to thank Professor Sudhirkumar Barai of the Department of Civil Engineering, IIT Kharagpur for his continued and unconditional guidance. An exemplary teacher and a magnificent person, we consider ourselves lucky to have been taught by him and to have worked under his supervision. Without his course, \quotes{Soft Computing Tools in Engineering}, and his co-operation the preparation of this paper would not have been possible.

\ifCLASSOPTIONcaptionsoff
  \newpage
\fi



%

\bibliography{research}{}

\bibliographystyle{ieeetr}

%




\end{document}